\documentclass{article}

\usepackage[nonatbib, final]{tccml_neurips_2020}

\usepackage[utf8]{inputenc} 
\usepackage[T1]{fontenc}    
\usepackage{hyperref}       
\usepackage{url}            
\usepackage{booktabs}       
\usepackage{amsfonts}       
\usepackage{nicefrac}       
\usepackage{microtype}      
\usepackage{amsmath}
\usepackage{amssymb}
\usepackage{wrapfig}
\usepackage{floatrow}
\usepackage{graphicx}
\usepackage{float}
\usepackage{xcolor}
\usepackage{enumitem}

\def\X{{\mathbf{X}}}
\def\x{{\mathbf{x}}}

\def\P{{\mathbb P}}

\def\methodname{\texttt{HECT}}

\newif\ifdraft
\drafttrue 

\usepackage[normalem]{ulem}
\newcommand\remove{\bgroup\markoverwith{\textcolor{orange}{\rule[.5ex]{2pt}{1pt}}}\ULon}

\title{HECT: High-Dimensional Ensemble Consistency Testing for Climate Models}

\author{%
  Niccol\`o Dalmasso\thanks{Equal Contribution} $^{\, ,1}$, Galen Vincent$^{*, 1}$, Dorit Hammerling$^{2}$ and Ann B. Lee$^{1}$ \\
  $^1$Department of Statistics \& Data Science, Carnegie Mellon University\\
  $^2$Department of Applied Mathematics and Statistics, Colorado School of Mines
}

\begin{document}

\maketitle
\vspace{-0.25cm}
\begin{abstract}
Climate models play a crucial role in understanding the effect of environmental and man-made changes on climate to help mitigate climate risks and inform governmental decisions.
Large global climate models such as the Community Earth System Model (CESM), developed by the National Center for Atmospheric Research, are very complex with millions of lines of code describing interactions of the atmosphere, land, oceans, and ice, among other components. 
As development of the CESM is constantly ongoing, simulation outputs need to be continuously controlled for quality. To be able to distinguish a ``climate-changing'' modification of the code base from a true climate-changing physical process or intervention, there needs to be a principled way of assessing statistical reproducibility that can handle both spatial and temporal high-dimensional simulation outputs.
Our proposed work uses probabilistic classifiers like tree-based algorithms and deep neural networks to perform a statistically rigorous goodness-of-fit test of high-dimensional spatio-temporal data.

\end{abstract}

\section{Introduction}
Climate change adaptation has been identified as one of the two main action items in addressing climate change \cite{rolnick2019climatechange}. The main goals in climate change adaptation are to anticipate the consequences of extreme climate events and understand the effects on climate of environmental and man-made changes. In this context, forecasts from climate models are essential for informing adaptation strategies, both at a local and national level (such as the IPCC reports \cite{ipcc2014mitigationofclimatechange, ipcc2018specialreport}). Also, climate models can be used to estimate the impact on the climate of a specific action or policy. In a recent example, scientists managed to estimate the effect of COVID19-related lockdown and travel restrictions on global warming (which turned out to be negligible) using the FaIR v1.5 carbon cycle emulator \cite{smith2018fair}.

Overall, climate models are divided into two categories, General Circulation Models (GCMs) or Earth System Models (ESMs) \cite{hugues2009introclimate}. Within the latter category, the Community Earth System Model (CESM \cite{hurrel2013CESM}) is a state-of-the-art ``virtual laboratory'' for studying past, present, and future global climate states. Developed mainly at the National Center for Atmospheric Research (NCAR), CESM consists of different simulation components covering various aspects of the climate system, including atmosphere, land, river runoff, land-ice, oceans, and sea-ice. CESM is a fully coupled system, meaning that the differential equations within each component are all solved at the same time by a custom computing architecture \cite{craig2012coupler}. About 2,900 academic publications have used CESM and CESM data, and the code base currently contains around 1.5 million lines of code.\footnote{Source: \url{http://www.cesm.ucar.edu/} and Hammerling et al., 2020 Joint Statistical Meeting}

\begin{figure}
    \centering
    \includegraphics[width=0.765\textwidth]{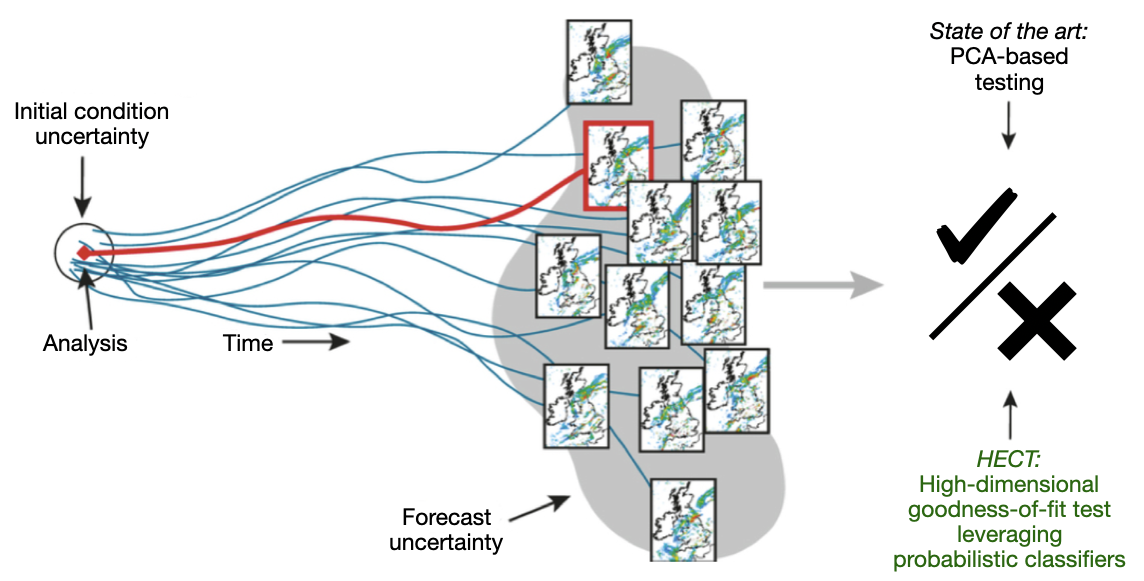}
    \caption{Schematics of the ECT methodology for quality assurance of climate models. A series of simulations with different initial conditions are run on a ``trusted'' machine (blue lines) to form a baseline ensemble which defines the forecast uncertainty (grey area). Test runs (red line) are then evaluated against this initial condition ensemble. The current state-of-the-art is to compress the data first by PCA, which may lead to estimation bias and loss of power. In this work, we propose a new testing framework (\texttt{HECT}) based on probabilistic classifiers that provides diagnostics and allows the data scientist to work directly with high-dimensional spatial and temporal outputs.}
    \label{fig:uncertainty_cesm}
\end{figure}

Such high-capacity and complex climate models are in a constant state of development, and frequent checks need to be in place for ``quality assurance'' \cite{oberkampf2010verificationandvalidation}; that is, to detect and reduce errors which could adversely affect the simulation results and potentially erode the model's scientific credibility. CESM is a deterministic simulation model, i.e., the output from two simulations are bit-for-bit (BFB) identical if they are run with the same initial conditions and on the same computing infrastructure. However, because of the chaotic nature of the equations underlying the simulation process, running the same code on new machine architecture, or running a tweaked version of the original code expected to return the same results, could produce output that is not BFB identical but still represents the same climate model. How can we make sure that non-BFB identical simulation outputs are a result of this natural variation rather than a ``climate-changing'' error we introduced in the code? \textit{Ensemble consistency testing} (ECT, \cite{baker2015ECT, milroy2016ECT, milroy2018ECT}) provides a statistical answer to such question. Figure~\ref{fig:uncertainty_cesm} provides a schematic representation of ECT. Multiple CESM runs are initialized using a ``trusted'' version of the climate model (in blue), where in each run the initial temperatures are perturbed on the order of $O(10^{-14})$. The ensemble of outputs from these slightly perturbed runs provides a baseline which approximates the natural variability in accepted simulation output (represented by the grey area). This ensemble is contrasted to outputs from a series of ``test runs'', i.e. runs from the modified version of the climate model (in red), and a pass or fail is issued based on the first 50 sample PCA scores for global climate variable averages at the last time step of the simulation. Currently, ECT is implemented as an automatic Python tool as part of the official CESM release. 


While PCA-based testing has proven successful, its main limitations are (i) a loss of statistical power from data compression and (ii) lack of theoretical guarantees on type I and II error. For (i), PCA-based testing only uses a subset of the PCA scores, compressing the data even further than averaging across the spatial and temporal resolution of simulation outputs. This loss of information makes the test more likely to fail in flagging truly ``climate-changing'' modifications (loss of statistical power), as well as biasing the PCA scores estimates (which are ultimately used to issue a pass/fail). In this paper, we propose a novel \textit{High-dimensional Ensemble Consistency Testing} (\methodname) approach, which aims to adress the limitations mentioned above. \methodname\ is based on a two-sample test \cite{kim20192stfull, dalmasso2020validationfull}, which leverages probabilistic classifiers and hence does not require a prior dimension reduction step, global averaging of climate variables, or limiting the test to the last time step; see Section~\ref{sec:methods} for specific examples of its applicability. In addition, the performance of the probabilistic classifier is shown to directly connect to the type I and type II error of the test \cite{dalmasso2020validationfull}, making it straightforward to select which classifier to use in each setting. Finally, \texttt{HECT} provides diagnostics by identifying statistically significant spatial and/or temporal differences between trusted and test runs. While our proposal is specific to CESM, we envision our methodology to apply to other climate models as well.


\section{Data} \label{sec:data} 
In \cite{milroy2018ECT}, which represents the latest development in CESM ECT, the trusted run ensemble is composed of 350 high-resolution runs, while there are much fewer test runs (usually less than five). Each run simulates 134 climate variables across the globe (approximated in 3-D by a $1^\circ$ square grid over $30$ vertical levels) for each time step of the simulation (which \cite{milroy2018ECT} set to 9 as a result of an analysis on ensemble variability). In practice, each run has approximately 1.7 billion entries,\footnote{$46,802$ grid cells across $30$ vertical levels for $134$ variables over $9$ time steps.} which cannot be used for comparison as is. The ECT in \cite{milroy2018ECT} is performed by averaging all variables globally, only considering the last time step and removing variables that are redundant, have zero variance, or do not show enough variability over all time steps. For the remainder of this work we will refer to a run as $\X$, but implicitly we assume some level of spatial averaging, temporal averaging or feature selection.

\section{ECT and Diagnostics via High-Dimensional Goodness-of-Fit Testing} \label{sec:methods}
Let $P_0$ and $P_1$ represent the distributions of trusted and test runs
respectively, and suppose $\mathcal{S}_0 = \{\X^0_1,\ldots,\X^0_m\} \overset{i.i.d.}{\thicksim} P_0$ and $\mathcal{S}_1 =\{\X^1_1,\ldots,\X^1_n\} \overset{i.i.d.}{\thicksim} P_1$ where typically $m \gg n$. To test the hypothesis $H_0: P_1=P_0$ versus $H_1: P_1 \neq P_0$, we introduce a binary random variable or class label $Y$, and interpret {$P_i$,  $i=0,1$,} as the class-conditional distributions of $\X$ given {$Y=i$}. The two-sample problem then becomes equivalent to testing for  independence between $\X$ and $Y$, i.e., $H_0: P_1= P_0$ is equivalent to 
 $H_0: \P(Y=1 |\X=\x) = \P(Y=1) \ \ \forall~\x \in \mathcal{X}$.
Motivated by this, 
\cite{kim20192stfull} propose a two-sample test based on the test statistic \begin{equation}\label{eq:global_teststats}
    \widehat{\mathcal T} = \frac{1}{n+m} \sum_{i=1}^{n+m} \left( \widehat{r}(\X_i) - \widehat{\pi}_1 \right)^2,
\end{equation}
where  $\widehat{r}(\x)$ is an estimate of $\P(Y=1|\X=\x)$ based on  $\{(\X_i,Y_i)\}_{i=1}^{n+m}$, and $\widehat{\pi}_1 =  \frac{1}{n+m} \sum_{i=1}^{n+m} I(Y_i=1)$ is an estimate of $\P(Y=1)$.
By the above reformulation, we have converted two-sample testing 
to the well-studied problem of estimating the ``class posterior'' $\P(Y=1|\X=\x)$, where we have access to an arsenal of high-dimensional regression (or probabilistic classification) algorithms for computing $\widehat{r}(\x)$. In settings where $m \gg n$ (i.e., there are many more trusted than test runs), such as in the CESM ECT example, we may gain power by using the test statistic in equation~\eqref{eq:global_teststats} in a goodness-of-fit test rather than in a two-sample test; for details see \cite{dalmasso2020validationfull} (Algorithm 5, Section D). In a nutshell, we generate several different ensemble runs $E$ of size $m_e$ to produce a set of realized test statistics $\{\widehat{T}^{(e)}\}_{e=1}^{E}$; such set is then used to define a null distribution for testing the hypothesis that trusted and test runs are statistically indistinguishable. 

\textbf{Contribution.} The main benefit of \methodname\ is that one now can eliminate most of the intermediate steps between simulation output and ECT (spatial averaging, global averaging, variable selection and dimensionality reduction). Such bottlenecks could potentially be avoided by leveraging probabilistic classifiers within \methodname\ which can handle high-dimensional spatial and temporal data. For example:
\begin{itemize}[leftmargin=*]
    \item We could leverage tree-based algorithms (e.g, Random Forest~\cite{breiman2001randomforest} and Gradient Boosted trees~\cite{friedman2001boostingmachines}), which are robust to highly correlated features and implicitly perform feature selection;
    \item Grid-level global data can be used as input to to convolutional neural networks~\cite{lecun1998cnn, alexnet2012cnn, he2016resnetcnn} without a prior dimension reduction step, thereby detecting local differences between trusted and test runs otherwise masked by global averaging input;
    \item Similarly, multivariate time-series can be directly input into suitable classifiers such as recurrent neural networks~\cite{Elman90findingstructure} and related recent developments~\cite{yu2017timeseriestensor, qin2017attentiontimeseries, rangapuram2018deepstatespace, lai2018lstnet}, thereby taking the entire simulated sequence of climate variables into account instead of only the last time step as in~\cite{milroy2018ECT};
    \item Finally, we could potentially leverage spatio-temporal deep neural networks~\cite{karpathy2014videodetection, shi2015convlstm} to compare runs that are only averaged over the vertical level of the atmosphere, hence analyzing 
    information at a global level over all time steps of the simulation.
\end{itemize}

For all settings above, we plan on exploring deep transfer learning techniques to aid training in cases where a limited amount of simulation runs is available (see \cite{tan2018transferlearning} for a review).
Corollary 2 of Theorem 2 in \cite{dalmasso2020validationfull} provides a connection between the mean integrated squared error (MISE) and the type I and type II error of the proposed \methodname, with no restrictions on the classifier form (i.e., also including neural networks). What this means in practice is that we now have a concrete way of deciding which regression or probabilistic classification algorithm to implement in different settings based on (anticipated) highest statistical power. In addition, the test statistic~\eqref{eq:global_teststats} provides {\em diagnostics} by identifying statistically significant differences (if any) between runs in a potentially high-dimensional spatio-temporal feature space; \cite{kim20192stfull} showcase the capability of diagnostics in feature space by comparing multivariate distributions of galaxy morphology images.

\textbf{Planned Experiments and Evaluation Metrics.} 
We plan on first using the ensemble runs of \cite{milroy2018ECT} (taking the global averaged at the 9$^{\rm th}$ time step for each run) and reproduce the common experiments among \cite{milroy2018ECT, milroy2016ECT, baker2015ECT} to explore whether \methodname\ issues a pass or fail in agreement with current ECT methodologies. We also propose to look at the performance of \methodname\ when we include more time steps and/or global-level data (i.e., averaged only on the vertical levels). Finally, we will determine the optimal ensemble size for different settings of \methodname, in similar fashion as what was done by \cite{milroy2018ECT}, Section 4.

\section*{Acknowledgments}
This work is supported in part by the NSF AI Planning Institute for Data-Driven Discovery in Physics, NSF PHY-2020295.

\end{document}